\documentclass[12pt,english]{article}
\pdfoutput=1
\usepackage{setspace}
\usepackage{dcolumn}
\usepackage{bm}
\usepackage{amsmath, amssymb, amsthm, mathrsfs}
\usepackage{graphicx}
\usepackage{listings}
\usepackage[T1]{fontenc}
\usepackage{cite}

\usepackage[unicode=true,bookmarks=true,
bookmarksnumbered=false,bookmarksopen=false,
breaklinks=false,
pdfborder={0 0 1},
backref=false,colorlinks=true]{hyperref}

\hypersetup{pdftitle={Branches of the Black Hole Wave Function Need Not Contain Firewalls}, pdfauthor={Ning Bao, Sean M. Carroll, Aidan Chatwin-Davies, Jason Pollack, Grant N. Remmen}, citecolor=black,linkcolor=black,urlcolor=black}

\newcommand{\Eq}[1]{Eq.~(\ref{#1})}
\newcommand{\Eqs}[2]{Eqs.~(\ref{#1}) and (\ref{#2})}
\newcommand{\Sec}[1]{Sec.~\ref{#1}}

\newcommand{\Fig}[1]{Fig.~\ref{#1}}

\newcommand{\Ref}[1]{Ref.~\cite{#1}}
\newcommand{\Refs}[1]{Refs.~\cite{#1}}
\newcommand{\Hil}{\mathcal{H}}
\newcommand{\mrm}[1]{\mathrm{#1}}
\newcommand{\scri}{\mathscr{I}}

\newcommand{\ket}[1]{| #1 \rangle}

\newcommand{\ketbra}[2]{|#1\rangle \langle #2 |}

\newcommand{\be}{\begin{equation}}
\newcommand{\ee}{\end{equation}}

\DeclareMathOperator{\Tr}{Tr}

\usepackage{color}
\definecolor{purple}{rgb}{0.5,0,0.5}
\definecolor{orange}{rgb}{0.7,0.2,0}

\def\simgt{\mathrel{\lower2.5pt\vbox{\lineskip=0pt\baselineskip=0pt
           \hbox{$>$}\hbox{$\sim$}}}}
\def\simlt{\mathrel{\lower2.5pt\vbox{\lineskip=0pt\baselineskip=0pt
           \hbox{$<$}\hbox{$\sim$}}}}

\begin{document}

\baselineskip=18pt
\hfill CALT-TH-2017-068
\hfill

\vspace{1cm}
\thispagestyle{empty}
\begin{center}
{\LARGE\bf
Branches of the Black Hole Wave Function\\Need Not Contain Firewalls
}\\
\bigskip\vspace{0.5cm}{
{\large Ning Bao,\textsuperscript{1} Sean M. Carroll,\textsuperscript{2} Aidan Chatwin-Davies,\textsuperscript{2} \\Jason Pollack,\textsuperscript{3} and Grant N. Remmen\textsuperscript{1}\\~}
} \\
 {\it \textsuperscript{1}\mbox{Berkeley Center for Theoretical Physics,} \\ \mbox{University of California, Berkeley, CA 94720, USA}\vspace{1mm}
\textsuperscript{2}\mbox{Walter Burke Institute for Theoretical Physics,}\\\mbox{California Institute of Technology, Pasadena, California 91125, USA}\vspace{1mm}
\textsuperscript{3}\mbox{Department of Physics and Astronomy,}\\ \mbox{University of British Columbia, Vancouver, BC, V6T 1Z1, Canada}\\ ~
}
\let\thefootnote\relax\footnote{\hspace{-7.5mm} e-mail:\\ \href{mailto:ningbao75@gmail.com}{\tt ningbao75@gmail.com},
\href{mailto:seancarroll@gmail.com}{\tt seancarroll@gmail.com},
\href{mailto:achatwin@caltech.edu}{\tt achatwin@caltech.edu}, \\
\href{mailto:jpollack@phas.ubc.ca}{\tt jpollack@phas.ubc.ca},
\href{mailto:grant.remmen@berkeley.edu}{\tt grant.remmen@berkeley.edu}} \\
 \end{center}
\bigskip
\centerline{\large\bf Abstract}

\begin{quote} \small
We discuss the branching structure of the quantum-gravitational wave function that describes the evaporation of a black hole. 
A global wave function which initially describes a classical Schwarzschild geometry is continually decohered into distinct semiclassical branches by the emission of Hawking radiation. 
The laws of quantum mechanics dictate that the wave function evolves unitarily, but this unitary evolution is only manifest when considering the global description of the wave function; it is not implemented by time evolution on a single semiclassical branch. 
Conversely, geometric notions like the position or smoothness of a horizon only make sense on the level of individual branches. 
We consider the implications of this picture for probes of black holes by classical observers in definite geometries, like those involved in the AMPS construction.
We argue that individual branches can describe semiclassical geometries free of firewalls, even as the global wave function evolves unitarily.
We show that the pointer states of infalling detectors that are robust under Hamiltonian evolution are distinct from, and incompatible with, those of exterior detectors stationary with respect to the black hole horizon, in the sense that the pointer bases are related to each other via nontrivial transformations that mix the system, apparatus, and environment. 
This result describes a Hilbert-space version of black hole complementarity.
\end{quote}
	
\setcounter{footnote}{0}

\newpage
\tableofcontents
\newpage

\section{The black hole information puzzle}

In 1975, Stephen Hawking showed that, in coordinates stationary with respect to a static black hole, quantum fields outside the black hole horizon are in a state of outgoing radiation that is very nearly thermal \cite{Hawking:1974sw}.
The backreaction of this thermal emission should lead even astrophysical black holes to evaporate over time, gradually transferring their mass into an ensemble of dilute radiation.
However, upon extrapolating Hawking's result to the case of a completely evaporating black hole, one is confronted with an apparent departure from quantum mechanics: it appears that when a pure state of matter---and the quantum information that it encodes---collapses into a black hole that then evaporates, it has evolved into a thermal mixed state and lost its coherent information.
Whether and how the quantum state can evolve unitarily from before a black hole is formed to after it evaporates
is known as the black hole information puzzle \cite{Hawking:1974sw, Susskind:1993if, Polchinski:2016hrw, Wallace:2017wzs}.

Several renditions of the black hole information puzzle have emerged over the last few decades.
In its modern form, the information puzzle is neatly summarized as a conflict between the following four postulates, articulated by Almheiri \emph{et al.} (AMPS) \cite{Almheiri:2012rt}:
\begin{enumerate}
\item \emph{Unitarity} --- As viewed by an observer who remains far away from the black hole, the formation and evaporation of the hole is a unitary quantum-mechanical process.
\item \emph{Local Effective Field Theory} --- To the exterior of the black hole's stretched horizon \cite{Susskind:1993if,Thorne:1986iy}, the physics of matter is well described by a local effective field theory on a black hole spacetime background.
\item $S_{\rm bh} = S_{\rm BH}$ --- As viewed by an observer who remains far away from the horizon, the black hole is a quantum-mechanical system that is represented by a finite-dimensional Hilbert space.
Moreover, the von~Neumann entropy of an old black hole, $S_{\rm bh}$, is (if not exactly, approximately) equal to its Bekenstein-Hawking entropy, $S_{\rm BH}$.
\item \emph{No Drama} --- An observer who crosses the apparent horizon of the black hole (but remains far from its central singularity) encounters nothing that runs contrary to the predictions of semiclassical general relativity and effective field theory.
\end{enumerate}

Taken together, these postulates seemingly imply a violation of monogamy of entanglement \cite{PhysRevLett.30.434}. 
This is because, while the second and fourth postulates together imply maximal entanglement between a portion of the the black hole interior and the late Hawking radiation, the first and third together imply that the late Hawking radiation must purify the early radiation as it is emitted.
These constraints on the entanglement shared among the black hole, the early Hawking radiation, and the late Hawking radiation cannot be mutually satisfied without violating strong subadditivity of entanglement entropy.
It would seem that taking all four postulates to be true leads to a contradiction, which must be resolved by requiring at least one of them to be violated in practice.

Several different resolutions to this puzzle have been proposed,\footnote{This list is not meant to be exhaustive---for one listing see the comprehensive bibliography in \Ref{Almheiri:2013hfa}.} from those that modify quantum mechanics \cite{Papadodimas:2012aq, Lloyd:2013bza} to those that allow a breakdown of no drama \cite{Almheiri:2012rt} or of unitarity \cite{Unruh:2017uaw}, identify the early Hawking radiation with the black hole interior \cite{Maldacena:2013xja}, modify the interior geometry \cite{Mathur:2005zp,Nomura:2014woa,Hertog:2017vod}, invoke quantum complexity theory \cite{Harlow:2013tf, Bao:2015hdp, Bao:2016uan}, allow for black hole remnants \cite{Chen:2014jwq}, or take nonlocal approaches \cite{Giddings:2012gc,Osuga:2016htn}.

By formulating black hole formation and evaporation as a process in Hilbert space in the context of Everettian quantum mechanics, we will argue that the four postulates above are made mutually consistent once we appreciate that the situations they refer to are not directly comparable.
In particular, a prerequisite for both local effective field theory and no drama is the presence of a semiclassical background geometry.
We will argue, as have several authors before us \cite{Hsu:2013cw, Hsu:2013fra,Hartle:2015bna,Hertog:2017vod}, that in a fully quantum-gravitational treatment, an evaporating black hole is described not by a single semiclassical background but rather a superposition of many such geometries, each corresponding to different branches of the wave function.\footnote{This conclusion could be viewed as a (mild and well-understood) violation of the second postulate above---there is not one local effective field theory for a single background but rather a different effective field theory on each semiclassical background. Because properties (such as the location) of the horizon can differ from branch to branch, our argument is reminiscent of state-dependent resolutions to the firewall paradox, e.g., Refs.~\cite{Papadodimas:2015jra,Nomura:2013gna,Verlinde:2013uja}. We emphasize that this state dependence arises naturally from the decoherence of the wave function and is not a violation of quantum mechanics but rather a consequence of the fact that geometric properties differ from branch to branch.}

In short, while unitarity applies to the global wave function, the no-drama condition only applies on branches of the wave function. 
Therefore, the AMPS construction \cite{Almheiri:2012rt} does not lead to a paradox, as its components do not  necessarily imply violation of monogamy of entanglement.
Similar points have previously been made schematically \cite{Nomura:2012sw, Nomura:2012ex, Hsu:2013cw, Hsu:2013fra,Hartle:2015bna,Hertog:2017vod,Yeom:2016qec}, but in this work we will give a more precise articulation of this view. 
In doing so, we will also find that---under the reasonable assumption that each Hawking quantum has the opportunity to interact with the rest of the universe after it has been emitted---the production of a large number of decohered branches allows Hilbert-space subfactors associated with the emitted radiation to have a large von Neumann entropy in the global wave function, even while they remain unentangled in every branch.
This sort of entropic structure allows firewall-free individual branches while preserving the Page curve as a statement about the global wave function.
Such a picture is heuristic at best, however, since the factorization of Hilbert space into early radiation, late radiation, and black hole degrees of freedom is highly branch dependent.

According to the principle of black hole complementarity \cite{Susskind:1993if}, we should not expect to be able to use local quantum field theory to simultaneously describe physics on both sides of a spacelike slice crossing an event horizon; what appear as local degrees of freedom inside a black hole will be distributed across the stretched horizon from the point of view of an external observer.
Following our Hilbert-space perspective, we argue that this principle can be implemented in terms of how Hilbert space is factorized into subsystems and what basis of pointer states is associated with the resulting decomposition.
The states that are robust with respect to environmental monitoring from the point of view of an infalling observer will appear fragile to an outside observer.
We exhibit an example decomposition of the relevant Hilbert spaces for each observer to show how this can work in practice.

Many puzzles about black hole evolution and evaporation certainly remain, such as whether the no-drama condition can be preserved at the level of the global wave function \cite{Bousso:2013wia}, how to reconstruct the black hole interior \cite{Papadodimas:2012aq}, and whether entanglement and wormholes are inextricably related \cite{Maldacena:2013xja}.
Moreover, determining whether firewalls or smooth horizons with no drama are typical requires an analysis of the detailed branching structure of the global wave function for an evaporating black hole. 

The rest of this paper is structured as follows.
We begin in \Sec{sec:unitarity} by carefully formulating the process of black hole formation and evaporation so that we may properly discuss unitarity in a fully quantum-gravitational sense.
Within this framework, we then investigate what it means to operationally probe entanglement between the black hole and exterior degrees of freedom in \Sec{sec:operational}.
We end with some brief concluding remarks in \Sec{sec:conclusion}.

\section{What is unitary and what is not}
\label{sec:unitarity}

\subsection{Setup}

To examine unitarity for black hole formation and evaporation, let us set up the problem as a scattering experiment, employing the S-matrix ansatz \cite{Giddings} for asymptotically flat spacetime. 
Suppose that the initial state is a pure state of dilute matter that will collapse to form a black hole, specified on the past boundary of an asymptotically flat spacetime,
\begin{equation}
\ket{\Psi_i} \equiv \ket{\Psi(i^- \cup \scri^-)}.
\end{equation}
We define the initial state on the asymptotic past boundary so that it can be thought of as effectively some free-field-theoretic state without gravitational interactions.\footnote{We will not consider the subtleties in the S-matrix formulation relating to infrared divergences; see Refs.~\cite{Giddings,Strominger} and references therein.}
If quantum gravity is unitary, then this state unitarily evolves to another pure state; according to the S-matrix ansatz, the final pure state is given as a superposition of states each defined on the future boundary of an asymptotically flat spacetime,
\begin{equation}
\ket{\Psi_f} \equiv S \ket{\Psi_i} = \sum_j S_{i\rightarrow j} \ket{\Psi_j(i^+ \cup \scri^+)}.
\end{equation}

\begin{figure}[h]
\centering
\includegraphics[scale=0.8]{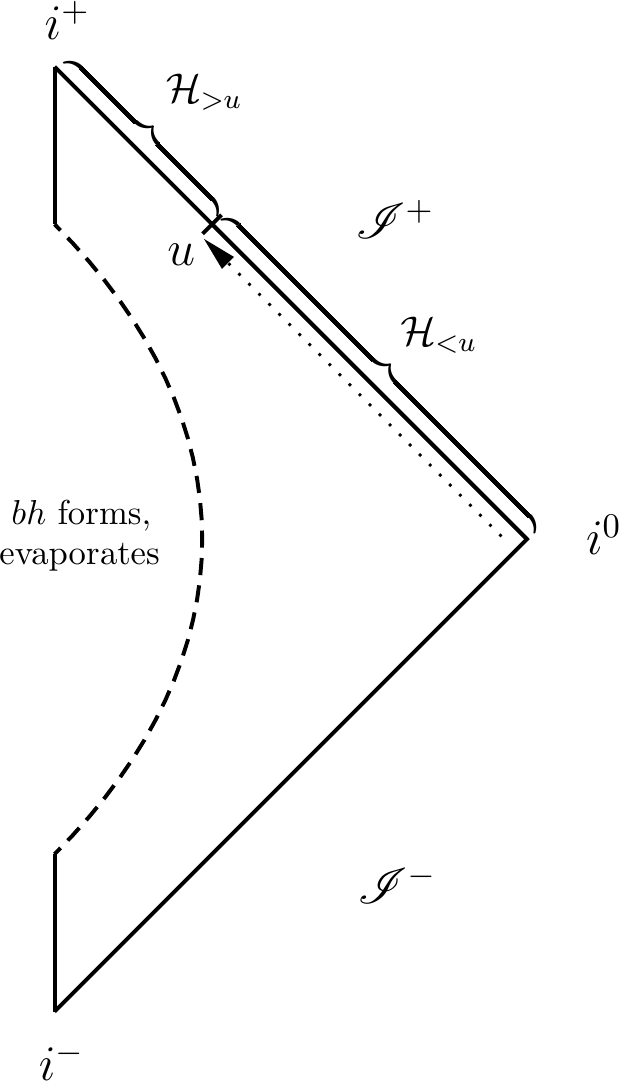}
\caption{The Penrose diagram for the spacetime that corresponds to a classical branch of the global wave function that itself describes the unitary formation and evaporation of a black hole in asymptotically flat spacetime. On the asymptotic future boundary, we divide the global Hilbert space into two factors, $\Hil_{<u}$ and $\Hil_{>u}$, the degrees of freedom of which lie to the past and future of the retarded time $u$, respectively. The direction of increasing $u$ is indicated by the dotted arrow. The asymptotic future $i^+ \cup \scri^+$ is identified across every classical branch of $\ket{\Psi}$ so that $\Hil_{<u}$ and $\Hil_{>u}$ are globally defined.}
\label{fig:HonestPenrose}
\end{figure}

Although the asymptotically flat spacetimes, each corresponding to a branch $j$, are not identical, by definition each of them has the same boundary geometry (with $\scri^+$ topology $S^{D-2} \times \mathbb{R}$). 
With an appropriate choice of coordinates, therefore, we can think of $\ket{\Psi_f}$ as a state that describes a superposition of definite field configurations on $i^+ \cup \scri^+$. 
In general this time evolution is not described by a single Penrose diagram, since, in the bulk, the quantum-gravitational evolution of the wave function does not correspond to a single classical geometry.\footnote{The most general S-matrix setup would describe a wave function defined on some number of copies of $i^- \cup \scri^-$ (only one for our choice of initial state $\ket{\Psi_i}$) that evolves to one defined on some number of copies of $i^+\cup\scri^+$, with no definitive spacetime structure in the interior.} 
Nevertheless, since the states at past and future null infinity are effectively noninteracting, we can identify all of these boundaries even in the absence of a well-defined bulk spacetime.
A Penrose diagram for each individual process $S_{i\rightarrow j}$, if it exists, should look somewhat like the diagram sketched in \Fig{fig:HonestPenrose}: an asymptotically flat spacetime with some intermediate evaporating black hole geometry, the details of which we cannot resolve without an explicit understanding of quantum gravity.\footnote{See also Fig.~5 of \Ref{Hartle:2015bna}, which had previously advanced a similar view.}

\subsection{The Page curve: late-time entanglement structure}
\label{sec:pagecurve}

Consider factorizing the state $\ket{\Psi_f}$ as follows.
Given a particular value of retarded time $u$ on $\scri^+$, with $u=+\infty$ corresponding to $i^+$, let us split the Hilbert space into the part with support to the past of $u$ and the part with support to the future of $u$,
\begin{equation}
\Hil_{i^+ \cup \scri^+} = \Hil_{<u} \otimes \Hil_{>u}. \label{eq:Hilscri}
\end{equation}
The reduced state of the ``early'' Hawking radiation is then given by tracing over $\Hil_{>u}$,
\begin{equation}
\rho_{<u} = \Tr_{>u} \ketbra{\Psi_f}{\Psi_f},
\end{equation}
and the ``Page curve'' \cite{Page:1993wv}\footnote{See \Ref{Harlow:2014yka} for further discussion of the Page curve.} is the plot of the von~Neumann entropy of $\rho_{<u}$ as a function of $u$, which decreases to zero as $u$ grows to cover all of $i^+\cup \scri^+$,
\begin{equation}
S(\rho_{<u})|_{u=+\infty} = 0.
\end{equation}
That $S(\rho_{<u })$ vanishes when $u = + \infty$ is simply a consequence of unitary evolution, since the final state $\ket{\Psi_f}$ is correspondingly pure.\footnote{Maudlin \cite{Maudlin:2017lye} has recently emphasized that global unitary evolution is in principle consistent with information loss outside the black hole, since one can define disconnected Cauchy surfaces with respect to which the black hole interior persists as an effective ``baby universe.'' We do not consider this possibility here, as it would violate Postulate 3, $S_{\rm bh} = S_{\rm BH}$. See also \Ref{Wallace:2017wzs}.}
In other words, in the global wave function, the ``late'' Hawking radiation purifies the ``early'' radiation.\footnote{We could have considered a spacetime with a timelike boundary, e.g., an asymptotically anti-de Sitter spacetime, but in that case defining the S-matrix proves difficult, for reasons discussed in for example \Ref{Fitzpatrick:2011dm}.}

We define the Page curve in terms of portions of the asymptotic future boundary because this definition does not rely on any particular choice of basis (for example, wave packets) for the Hawking radiation.
We also remain agnostic about the exact shape of the Page curve resulting from this division of the final state into early and late radiation.
Nevertheless, it is certainly true that $S(\rho_{<u})$ vanishes when $\rho_{<u}$ has support either nowhere or everywhere on the asymptotic future boundary.

\subsection{Unitary evolution, branches, and decoherence}

The modern black hole information problem arises when trying to interpret the entanglement structure at earlier times.
Previously, we have only discussed the initial-state and late-time structure of the global wave function. 
However, because in this paper we are assuming that the (as yet unknown) theory of quantum gravity is a bona fide quantum-mechanical theory, we can also write down the wave function at intermediate times.
Hence, the evolution of the state is, as usual, governed by the Schr\"{o}dinger equation:\footnote{In canonical quantum gravity, we could also take the point of view that the wave function should obey the Wheeler-DeWitt equation \cite{1967PhRv..160.1113D}.
In this case, $\hat H$ is the Hamiltonian constraint, $\hat H \ket{\Psi} =0$, and we need some additional information to implement time evolution as an emergent phenomenon. This approach is also proposed in \Refs{Hartle:2015bna,Hertog:2017vod}.}
\be
\hat H \ket{\Psi} = i \frac{d}{d\lambda} \ket{\Psi}.
\ee
We emphasize that this equation genuinely implements time evolution; however, because $\lambda$ need not have any relation to any coordinate or proper time in a geometric description,\footnote{In a holographic description, we could think of $\lambda$ as the time coordinate of the boundary theory.} we have chosen to use $\lambda$ rather than $t$. 
Implementing our chosen boundary conditions, we must have $\ket{\Psi(0)} = \ket{\Psi_i}$ and $\ket{\Psi(1)} = \ket{\Psi_f}$.
Because its evolution is governed by the Schr\"{o}dinger equation, $\ket{\Psi}$ manifestly evolves unitarily.

A challenge in interpreting the state $\ket{\Psi(\lambda)}$ at intermediate values of $\lambda$ lies in the fact that it does not describe a single black hole geometry.
That is, a single geometry at one time (for example, $\lambda = 0$) must evolve to a state that describes an ensemble of many possible geometries at a later time.
An observer or detector present in the initial state would see different measurement outcomes depending on what geometry they were in at a later time.

For instance, while the expectation value of the black hole position and momentum remains fixed and constant in the global wave function, an observer who is monitoring the black hole would measure a drift in its position and momentum as it receives kicks from Hawking quanta that are emitted and interact with the surrounding environment, leading to decoherence.
In Everettian language, the notion of a classical black hole geometry exists only on decohered branches of the global wave function.
Therefore, in order to have an idea of a definite geometry throughout black hole formation and evaporation, it is necessary to specify what the decohered branches of the wave function are and what determines this branching structure.

This leads us to conjecture that the emergence of an ensemble of classical geometries from the unitary evolution of $\ket{\Psi(\lambda)}$ can be understood as a decoherence process, which determines a pointer basis for $\ket{\Psi(\lambda)}$ of which the elements describe decohered geometries and configurations of matter. 
The lesson of the decoherence program \cite{Zeh:1970fop,Zurek:1981xq,Griffiths:1984rx,Joos:1984uk,Schlosshauer:2003zy} is that branching of the wave function is set by the interaction dynamics between a particular subsystem and the environment monitoring this subsystem. 
In order to determine the branching structure, we need to decompose the Hilbert space into ``system'' and ``environment'' degrees of freedom.
For our purposes, we suppose that there exists a set of degrees of freedom in the total Hilbert space that can serve as an environment that, minimally, yields a definite geometry when traced over.
For example, one could conjecture this environment to comprise some inherently quantum-gravitational degrees of freedom.\footnote{For some discussion of this kind of UV/IR factorization, see \Refs{Jacobson:2015hqa, Harlow:2016vwg, Cao:2017hrv}.}
Alternatively, recent studies suggest that the modes of soft gravitons and other soft massless gauge bosons may constitute such an environment \cite{Carney:2017jut,Carney:2017oxp,Carney:2018ygh,Strominger:2014pwa,Hawking:2016msc,Strominger:2017aeh}.
Regardless, we conjecture that the global Hilbert space may be written as 
\begin{equation}
\Hil = \Hil_\text{eff} \otimes \Hil_\mrm{env}, 
\end{equation}
so that the global state takes the form of a sum over branches,
\begin{equation}
\ket{\Psi(\lambda)} = \sum_b \alpha_b(\lambda) \ket{\Psi_b} ,\label{eq:global}
\end{equation}
where each branch $\ket{\Psi_b}$ decomposes as
\begin{equation}
\ket{\Psi_b} = \ket{\psi_b}_\text{eff} \otimes \ket{\varepsilon_b}_\mrm{env} \, .
\label{product-state-structure}
\end{equation}
Then $\ket{\psi_b}_\text{eff}\in \Hil_{\rm eff}$ is the part of the state that we may think of as describing a semiclassical geometry and the states of quantum fields in the theory on top of this geometry, while $\ket{\varepsilon_b}_\mrm{env}$ is a state of the environmental degrees of freedom that are responsible for decohering the state to a semiclassical geometry.
How the global wave function branches depends on how the $\ket{\varepsilon_b}_\mrm{env}$ are determined.

Depending on the superselection rules of quantum gravity, we might only need to consider, e.g., branches $b$ that correspond to asymptotically flat geometries, geometries with identical topologies to the initial state, etc.
We can either implement these rules by working in a smaller Hilbert space than the full Hilbert space of quantum gravity or by imposing that $\alpha_b(\lambda)=0$ for all branches $b$ corresponding to geometries that do not obey these superselection rules.\footnote{For example, thought experiments in AdS/CFT suggest the existence of topological superselection rules for holographic wormhole geometries \cite{Marolf:2012xe}.}

Trivially, the Hilbert space $\Hil_\mrm{eff}$ admits a direct sum structure \cite{Nomura:2011dt,Pollack:2018yum},
\begin{equation}
\Hil_\mrm{eff} = \bigoplus_b \mrm{span}\{ \ket{\psi_b}_\mrm{eff} \} .
\end{equation}
However, we expect that it should be possible to group together sets of states that have the same background geometry to form subspaces,
\begin{equation}
\Hil_\mrm{eff}^{\mathcal{B}} \equiv \mrm{span} \left\{ \ket{\psi_b}_\mrm{eff} ~ | ~ b \in \mathcal{B} \right\},
\end{equation}
where any given $\mathcal{B}$ contains the labels of a set of branches that all correspond to the same background geometry (to within some precision that specifies a coarse graining).
As such, we envision each $\Hil_\mrm{eff}^{\mathcal{B}}$ as being the Hilbert space of fields coupled to the background geometry of the branches in $\mathcal{B}$.
Each $\Hil_\mrm{eff}^{\mathcal{B}}$ can of course be further decomposed, for example, as a tensor product over the Hilbert spaces of the different species of fields contained in the effective theory.
Because field theories are defined on fixed spacetime backgrounds, a tensor product in each individual $\Hil_\mrm{eff}^{\mathcal{B}}$ does not necessarily extend to a tensor product on the entire semiclassical Hilbert space $\Hil_\mrm{eff}$.
Generally we can therefore write
\begin{equation}
\Hil_\mrm{eff} = \bigoplus_{\mathcal B} \Hil_\mrm{eff}^{\mathcal{B}} = \bigoplus_{\mathcal B} \left(\bigotimes_i \Hil_i^{(\mathcal B)}\right),
\end{equation}
where the $\Hil_i^{(\mathcal B)}$ represent factors defined on the Hilbert space of the specific background geometry $\mathcal B$.
In particular, notions such as ``modes of outgoing Hawking radiation near the horizon'' are only well defined on specific branches, not on the global wave function.

During the process of decoherence itself, the action of the Hamiltonian entangles system and environment states, and the entropy of the system density operator $\rho_\text{eff}= \Tr_\mrm{env}|\Psi\rangle\langle\Psi|$ increases.
After decoherence, $\rho_\text{eff}$ will be diagonal with respect to a basis of ``pointer states'' for $\Hil_\text{eff}$, each pointer state defining a different branch of the wave function. 
For us, the pointer states are the $\{\ket{\psi_b}\}$, representing quantum fields on a definite semiclassical background.
Once this occurs, branches interact minimally with each other (they decohere), so that the time evolution of a superposition of branches is approximately the same as evolving each branch individually.
In particular, the branches retain their product-state structure (\ref{product-state-structure}) under the action of the Hamiltonian implementing time evolution.

Returning to \Eq{eq:global}, we see that at each time $\lambda$ there are a number of decohered branches, describing a superposition of the geometries corresponding to those $\ket{\psi_b}_\text{eff}$ with $\alpha_b(\lambda) \ne 0$. 
As $\lambda$ increases, so does the number of decohered branches, i.e., the size of the set $\{\ket{\Psi_b} | \alpha_b(\lambda) \ne 0\}$.
It seems natural to relate this repeated branching to the production of entropy. As a result, the increase in the number of decohered branches is important for the interpretation of the Hawking entropy formula and the Page curve, as we discuss in 
\Sec{sub:branch_counting}.

\subsection{Entanglement structure at intermediate times}\label{sec:intermediate}

The basic reason why the consideration of branching structure is relevant to the black hole information puzzle is that, while evolution of the global wave function is unitary, evolution via conditioning on a specific background geometry (i.e., projection onto individual branches of the wave function) is not.
A particular sequence of classical states is produced by repeated nonunitary projection of the wave function onto states of definite background geometry; we refer colloquially to wave function ``collapse'' during the measurement process (which for us is simply decoherence and branching).
In particular, the Page curve, which we have seen above is a consequence of unitarity, only needs to hold for the global wave function.

Our main observation is that arguments for the modern information puzzle---and in particular Postulates 2 and 4 above---only apply at the level of the $\ket{\psi_b}_\text{eff}$ parts of the classical branches \cite{Hsu:2013cw, Hsu:2013fra}.
Again, evolution of the global wave function is unitary, but evolution at the level of individual classical geometries is not.
In \Sec{sec:tests}, we will discuss what it means to operationally probe the information puzzle in the context of this observation.
In essence, at intermediate times, it is not clear how to calculate the Page curve as we have formulated it in \Sec{sec:pagecurve} because the specification of what degrees of freedom constitute ``early'' radiation is a branch-dependent notion.

One possibility could be to make a branch-dependent local tensor product decomposition of the type suggested by AMPS, where ${\cal H}_{\rm eff}$ is taken to be $A^{(b)} \otimes B^{(b)} \otimes R^{(b)} \otimes C^{(b)}$,
where $R^{(b)}$ denotes degrees of freedom that correspond to early radiation, $A^{(b)}$ corresponds to the black hole degrees of freedom, $B^{(b)}$ is the late radiation, and $C^{(b)}$ (for ``complement'') is everything else.
However, such a decomposition is problematic for a number of reasons.
Generically, it will not be the case that $R^{(b)},\, B^{(b)},\, A^{(b)}$ are the same factors on every classical branch $\ket{\Psi_b}$.  
For instance, we can identify a space $A^{(b)}$ of black hole degrees of freedom on branches where a black hole exists and it is likely that this space may be consistently identified across all branches $b \in \mathcal{B}$ with the same background geometry.
But we cannot speak of anything like a global space of black hole degrees of freedom in the Hilbert space of all semiclassical states.
Moreover, even within branches that describe the same background geometry, Hilbert-space subfactors that consist of quanta of field excitations, such as $R^{(b)}$, will vary from branch to branch.
(For example, a branch with one decohered graviton of energy $E$ and a branch with two decohered gravitons of energies $E_1 + E_2 = E$ are distinct.)

Furthermore, a decomposition such as $A^{(b)} \otimes B^{(b)} \otimes R^{(b)} \otimes C^{(b)}$ becomes a highly ``observer-dependent'' refinement, in the sense that the Hilbert-space factors are neither dictated by the theory itself nor have a fixed spacetime interpretation.
For example, according to the description of an observer outside of a black hole on a branch $b$, states in $A^{(b)}$ describe states of the black hole's stretched horizon.
For such an observer, the interior of the black hole is not a geometric place, which runs counter (but complementary) to the description of an infalling observer if the horizon is transparent.
We will return to the question of complementarity and further decomposition of the classical branches in $\Hil_\mrm{eff}$ in \Sec{sec:operational}, but for now we stress that none of our arguments assumes that the exterior observer can assign any classical geometric interpretation to the black hole interior.

One of our main conclusions about black hole evolution in the global wave function is that unitary evolution and the no-drama condition are compatible in principle, even without violating monogamy of entanglement. Unitarity is a global concept; it applies only to the global wave function and not to individual semiclassical branches of particular geometry. On the other hand, the requirement of no drama is a statement about individual decohered branches describing such semiclassical geometries; in particular, it is a requirement that the state of the quantum fields near the black hole take a particular structure (corresponding to the vacuum) {\it on the branch}, i.e., within $\ket{\psi_b}_{\rm eff}$, for most of the branches.

Let us consider a toy model to illustrate that what parties appear entangled on decohered branches of a wave function can be very different  from the structure of entanglement entropies in the global wave function.
Consider, for example, four qubits labeled $A$, $B$, $C$, and $D$ in the state
\begin{equation} \label{eq:easyglobal}
\ket{\Psi}_{ABCD} = \tfrac{1}{\sqrt{2}} \left(\ket{00}_{AD} + \ket{11}_{AD}\right) \otimes \tfrac{1}{\sqrt{2}} \left( \ket{00}_{BC} + \ket{11}_{BC} \right).
\end{equation}
In this state and tensor decomposition, the pairs $AD$ and $BC$ are unentangled, while $A$ and $D$, as well as $B$ and $C$, are entangled.
However, suppose that we treat $CD$ as an environment and posit a Hamiltonian with an interaction term between $AB$ and $CD$ of the form
\begin{equation}
H_{\rm int} = \sum_{\pm} \mathcal{O}_{AB}^{(1),\pm} \otimes \ketbra{\phi^{\pm}}{\phi^{\pm}}_{CD} + \mathcal{O}_{AB}^{(2),\pm} \otimes \ketbra{\psi^{\pm}}{\psi^{\pm}}_{CD},
\end{equation}
where $\ket{\phi^\pm} = \tfrac{1}{\sqrt{2}} (\ket{00} \pm \ket{11})$ and $\ket{\psi^\pm} =  \tfrac{1}{\sqrt{2}} (\ket{01} \pm \ket{10})$ denote Bell states.
Then it follows that the branching structure of $\ket{\Psi}_{ABCD}$ (the environmental $CD$ parts of which commute with the interaction Hamiltonian) looks like
\begin{equation}
\begin{aligned}
\ket{\Psi}_{ABCD} &= \frac{1}{2}\left(\ket{\phi^+}_{AB} \otimes \ket{\phi^+}_{CD} + \ket{\phi^-}_{AB} \otimes \ket{\phi^-}_{CD}\right. \\&\;\;\;\;\;\;\left. + \ket{\psi^+}_{AB} \otimes \ket{\psi^+}_{CD} - \ket{\psi^-}_{AB} \otimes \ket{\psi^-}_{CD} \right).
\end{aligned}
\end{equation}
In other words, while $AD$ and $BC$ are unentangled in the global wave function, in the sense that $S(AD) = 0$, the ``system'' subfactors $A$ and $B$ are entangled on every branch.
Tracing out $CD$ to obtain a reduced density matrix for $AB$ would reveal a set of distinct branches, all of which would exhibit maximal entanglement between $A$ and $B$.

More generally, the production of a large number of orthogonal branches through decoherence can lead to large von~Neumann entropies for subsystems in the global wave function.
Heuristically, this is what we expect to happen for AMPS-like tensor product factors; $A^{(b)}$ and $B^{(b)}$ must be highly entangled for drama-free branches, yet decoherence can produce large von~Neumann entropies for the collections of $A^{(b)} B^{(b)}$ and $R^{(b)}$ on the branches, due to classical uncertainty.
This is only a heuristic picture, since there is no consistent identification of AMPS-like tensor product factors across all branches that may be used to compute the Page curve at intermediate times.
Nevertheless, it is interesting to assess just how much entropy is produced by branching, e.g., within a given sector $\mathcal{B}$, which we now discuss.

\subsection{Branch counting}\label{sub:branch_counting}

Consider a simple idealization, according to which $AB \subset U^\dagger(\lambda;1)[\Hil_{>u}]$ and $R \subset U^\dagger(\lambda;1)[\Hil_{<u}]$ actually are consistently identified as the same factors across all branches (even though, as discussed above, that is not precisely the case in our scenario).
Here, $U(\lambda_1; \lambda_2)$ is the unitary evolution operator that maps a state at parameter value $\lambda_1$ to the state at parameter value $\lambda_2 > \lambda_1$.
In other words, here, we explicitly hypothesize that the Hilbert-space decomposition ($A^{(b)} \otimes B^{(b)} \otimes R^{(b)} \otimes C^{(b)}$) holds globally across all branches, and we explore the resulting consequences.\footnote{Alternatively, we can think of the mental exercise discussed here as taking place within a collection of branches $\mathcal{B}$ that initially have the same geometry: We first project onto a collection of branches $\ket{\Psi_b}$, $b \in \mathcal{B}$, on which we make this decomposition of $\Hil_\text{eff}$ into $ABRC$ and we then study further evolution of entanglement.}

Tracing over $\Hil_\mrm{env}$ and $C$ in the global wave function, the reduced state on $ABR$ can take the form
\begin{equation} \label{eq:BLE_branches}
\Tr_{\mrm{env},C} \ketbra{\Psi(\lambda)}{\Psi(\lambda)} = \sum_b p_b(\lambda) \, \rho_b^{AB} \otimes \rho_b^{R},
\end{equation}
where, on each branch, $AB$ and $R$ are unentangled (even though they are correlated globally).
Such entanglement structure is required in order to avoid, for example, a firewall arising from broken entanglement across the $AB$ subsystems between the black hole and outgoing late radiation modes. That is, on each branch defining a classical spacetime geometry, we let the quantum fields take the vacuum configuration at the horizon, as required by Postulate 4.

Even though $AB$ and $R$ are unentangled on every branch, there is still nonzero von~Neumann entropy for $AB$ and $R$ globally.
Consider the reduced state on $AB$ alone,
\begin{equation}
\rho^{AB}(\lambda) = \sum_b p_b(\lambda) \rho_b^{AB}.
\end{equation}
The Holevo information \cite{Holevo98,Bao:2017guc} of $\rho^{AB}$ is given by
\begin{equation}
\chi(\rho^{AB}) = S(\rho^{AB}) - \sum_b p_b \, S(\rho_b^{AB}) \,
\end{equation}
and is an upper bound on the accessible information of $\rho^{AB}$ and its corresponding ensemble. More importantly for our purposes, it is bounded by the Shannon entropy, $- \sum_b p_b \log p_b$, with saturation occurring when each $\rho_b^{AB}$ has orthogonal support \cite{Holevo98}.
Moreover, $S(\rho^{AB})$ can be bounded from below by using the concavity of entanglement entropy. Putting these bounds together, we have
\begin{equation}
\sum_b p_b \, S(\rho_b^{AB}) \leq S(\rho^{AB}) \leq \sum_b p_b \, S(\rho_b^{AB}) - \sum_b p_b \log p_b \, .
\end{equation}

In particular, $S(\rho^{AB})$ can in fact be quite large.
For example, in the case where each $\rho_b^{AB}$ has orthogonal support, then $S(\rho^{AB}) \approx \log N$ if each $p_b \approx 1/N$, where $N$ is the number of branches (i.e., the sum over $b$ runs from 1 to $N$).
An old black hole of mass $M$ will have been emitting Hawking quanta of typical energy $\lesssim 1/M$, so greater than $O(M/(1/M)) = O(M^2)$ emissions will have occurred in the black hole's past.
If each emission branches the global wave function by a constant factor, then the scaling of $N$ goes as $e^{M^2}$. In order to specify a branch, we must choose not only the mass and momentum of the black hole itself but the entire exterior spacetime geometry, which, via backreaction, depends on the distribution of all the Hawking radiation between the black hole and $\scri^+$. 
It is therefore plausible that each Hawking emission indeed branches the global wave function, as long as the emitted quantum becomes entangled with the environment.\footnote{If Hawking quanta never become entangled with something that could be labeled ``an environment,'' branching would not occur. In that case, however, there is no sensible way to assign a semiclassical geometry to the state and it is not appropriate to speak of a black hole, much less a firewall.}
Had we only considered the macroscopic properties of the black hole itself, the number of branches would be much smaller~\cite{Almheiri:2013hfa}. However, as long as there is anything else in the universe aside from the black hole for a Hawking quantum to interact with on its way to infinity, it is reasonable to treat the backreaction of the Hawking quantum on the spacetime as decohering the wave function into states of definite geometry. Even with no exterior matter outside the black hole, it is conceivable that the gravitational interaction of Hawking quanta is itself enough to decohere the geometry; an exploration of this question and its possible connection to recent work on soft gravitons and their associated symmetries \cite{Strominger:2014pwa}  lies beyond the scope of the present work.
Note that, in our setup, $S(\rho^{AB})$ scales in the same way as the Bekenstein-Hawking entropy of the black hole, $S_{\rm BH} \sim M^2$, so it may be possible to recover Postulate 3, $S_{\rm bh} = S_{\rm BH}$, via the branching structure alone,\footnote{We note that a similar argument was made in the context of the fuzzball program in \Ref{Mathur:2009zs}.} but the details of the branch counting also lie beyond the scope of the present work. 
Such an analysis of the branching structure would be necessary in order to guarantee that no-drama states are indeed generic for a randomly selected black hole horizon in the global wave function; here we merely want to emphasize that such states are plausible.

\section{Operational tests of the information puzzle}\label{sec:tests}
\label{sec:operational}

We now turn to the question of how a pair of observers would practically implement the AMPS thought experiment \cite{Almheiri:2012rt} to probe the state of the black hole inside and outside the event horizon.
Our main concern is to understand this implementation in the context of unitarity of the global wave function.
In particular, we will argue that the branching structure of the global wave function is such that the state vectors that are robust under Hamiltonian evolution---the \emph{pointer states} into which the global wave function branches---are very different inside and outside the horizon.
Specifically, the pointer bases corresponding to measurements made by an interior, infalling observer and an external, static observer are related to each other via nontrivial transformations that manifest the complexity of black hole scrambling.
This means that it is impossible for both the infalling observer behind the horizon and the external observer to exist on the same semiclassical branch of the wave function at their level of coarse graining.\footnote{See \Ref{Nomura:2012sw} for a discussion of related ideas.}

Let us suppose that, as part of the initial asymptotic data, we specify that there are two detectors at $i^-$, $D_\mrm{inf}$ and $D_\mrm{st}$, corresponding to infalling and stationary observers and that each begins in some ready state $d_0$.
Assume that the detectors are local, can be switched on and off, and are identical in operation.
We can decompose the effective Hilbert space as
\begin{equation}
\Hil_\text{eff} = \widetilde\Hil_\text{eff} \otimes D_\mrm{inf} \otimes D_\mrm{st}.
\end{equation}
Specifically, we isolate the finite-dimensional Hilbert spaces $D_\mrm{inf}$ and $D_\mrm{st}$ that represent the detectors' internal degrees of freedom that ultimately couple to some local system to realize measurement.
We suppose that all of the detectors' other degrees of freedom, such as kinematic degrees of freedom like position and momentum, are a part of $\widetilde\Hil_\text{eff}$.
At intermediate parameter values $\lambda$ with the detectors switched off, we therefore write the global wave function as
\begin{equation}
\ket{\Psi(\lambda)} = \sum_b \alpha_b(\lambda) \ket{\tilde \psi_b}_\text{eff} \otimes \ket{d_0}_{D_\mrm{inf}} \otimes \ket{d_0}_{D_\mrm{st}} \otimes \ket{\varepsilon_b}_\mrm{env} \, .
\end{equation}

Our aim is to consider a situation in which one detector, $D_\mrm{st}$, remains stationary outside of a black hole and the other, $D_\mrm{inf}$, falls into the same black hole and to compare the measurements reported by the two detectors.
We therefore begin by projecting onto a branch of the wave function corresponding to a single spacetime so that the two detectors agree on the background geometry.
The object of interest is thus a particular branch $b_\star$ of the form $\ket{\tilde \psi_{b_\star}}_\text{eff} \otimes \ket{d_0}_{D_\mrm{inf}} \otimes \ket{d_0}_{D_\mrm{st}}$.
For convenience, we have temporarily dropped the $\Hil_\mrm{env}$ factor because it plays no role once we have projected onto a branch (keeping in mind that $\Hil_\mrm{env}$ is necessary for further evolution of the initial branch to be unitary).

Equipped with a notion of background spacetime, we can now attempt to interpret $\Hil_\text{eff}$ in terms of spacetime regions and in the context of measurements performed by the detectors on the branch $b_\star$.
Suppose that, on the branch in question, $D_\mrm{inf}$ falls into the black hole while $D_\mrm{st}$ remains outside.
Further suppose that, on this branch, at some moment, both detectors switch on and become entangled with the local degrees of freedom that they probe.
Let us define factors of $\Hil_{\rm eff}$ on a Cauchy surface chosen such that its intersection with the infalling detector's worldline occurs inside the black hole.

Consider first the following decomposition of $\Hil_\text{eff}$, appropriate from the point of view of the stationary detector:
\begin{align}
\nonumber \Hil_\text{eff} &= A \otimes S \otimes E \otimes D_\mrm{st} \\
&\equiv S \otimes D_\mrm{st} \otimes \mathcal{E}.
\label{eq:stdecomp}
\end{align}
Here, 
\begin{itemize}
\item $A$ is the black hole Hilbert space, which, in the spirit of complementarity, we suppose represents states of the stretched horizon.
\item $D_\mrm{st}$ is the Hilbert space of the stationary detector.
\item $S$ is the collection of local degrees of freedom that constitute the system that the stationary detector measures.
\item $E$ are any remaining exterior degrees of freedom.
Altogether, $\mathcal{E} \equiv A \otimes E$ is the environment for the stationary detector.
\end{itemize}

\begin{figure}[t]
\centering
\includegraphics[scale=1]{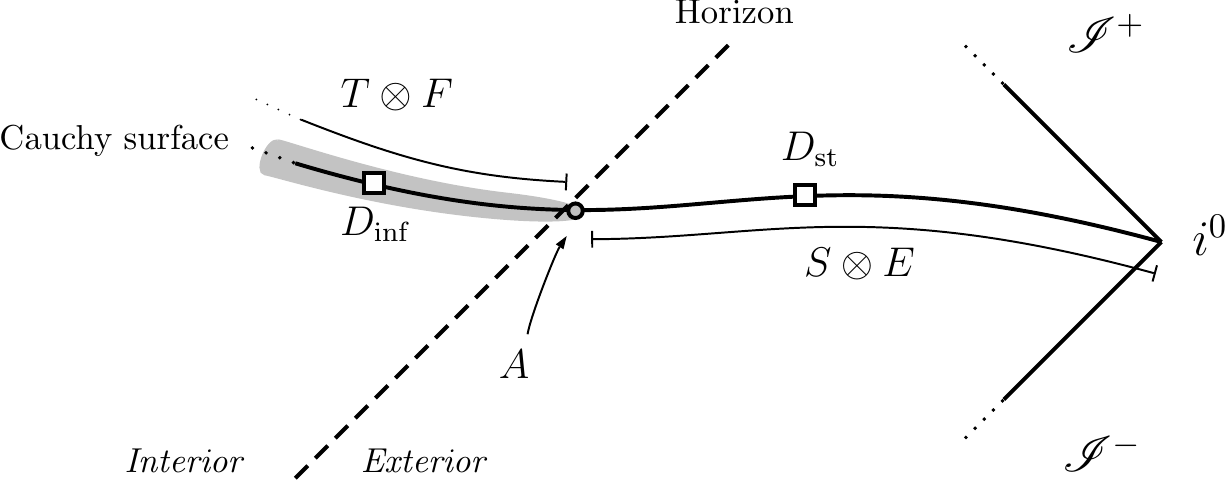}
\caption{Diagrammatic representation of the decompositions of $\Hil_\text{eff}$ in \Eqs{eq:stdecomp}{eq:infdecomp} on a Penrose diagram representing some particular semiclassical branch $b_\star$. The detectors and their associated internal Hilbert spaces, $D_\mrm{st}$ and $D_\mrm{inf}$, are denoted by the white boxes. The location of the stretched horizon and its associated Hilbert space, $A$, is denoted by the gray circle. According to black hole complementarity, we suppose that $A = D_\mrm{inf}\otimes T \otimes F$ are identified as the same Hilbert space. This is indicated by the shading of the part of the Cauchy surface in the black hole interior. The interior Hilbert-space factors and the interior geometry are only resolved by observers who cross the black hole's horizon. From the point of view of an exterior observer, these degrees of freedom are precisely the degrees of freedom of the stretched horizon. Also note that, according to \Eqs{eq:stdecomp}{eq:infdecomp}, $G = D_\mrm{st} \otimes S \otimes E$.}
\label{fig:decomp-new}
\end{figure}

Similarly, we can also decompose $\Hil_\text{eff}$ in a way that is appropriate for an infalling description:
\begin{align}
\nonumber \Hil_\text{eff} &= D_\mrm{inf} \otimes T \otimes F \otimes  G \\
&\equiv D_\mrm{inf} \otimes T \otimes \mathcal{F}.
\label{eq:infdecomp}
\end{align}
Here,
\begin{itemize}
\item $D_\mrm{inf}$ is the Hilbert space of the infalling detector.
\item $T$ is the system that the interior detector measures.
\item $F$ and $G$ are other degrees of freedom inside and outside the black hole, respectively.
Altogether, $\mathcal{F} \equiv F \otimes G$ is the environment for the stationary detector.
\end{itemize}
How the various Hilbert-space decompositions overlap is illustrated in Fig.~\ref{fig:decomp-new}.

One of the key results of black hole complementarity is that horizon dynamics, as seen by a stationary exterior observer, appear to be approximately typical with respect to the Haar measure on sufficiently long timescales.
This is discussed in, for example, \Refs{Sekino:2008he,Harlow:2013tf}.\footnote{\Ref{Thorne:1986iy} discusses classical black hole scrambling in the context of the membrane paradigm.} In the remainder of this section, we will find that black hole scrambling has important implications for the structure of the global wave function in terms of the pointer bases of interior and exterior observers.

Let us first develop some intuition for what to expect. Consider an infalling object crossing the stretched horizon as seen by either an observer falling along with the object or a stationary observer at some fixed position outside of the black hole.
While for the infalling observer the object will seem to pass through the horizon without any apparent effect, for the external observer the object will appear to scramble across and thermalize with the stretched horizon.
If the infalling object is a classical object---which in particular means that it is robust against decoherence due to monitoring by its environment in the infalling frame---this picture suggests that the object is explicitly not robust against decoherence due to environmental interactions \emph{in the frame of the external observer}, in which it is seen to quantum-mechanically scramble and delocalize across the entirety of the stretched horizon.
As the states that are robust against decoherence are by definition the pointer states, this highlights the fact that the pointer states in the infalling frame, when viewed in the frame of the static external observer, appear to be totally scrambled and delocalized.
In the context of black holes, this feature has traditionally been implemented by a unitary 2-design \cite{Hayden:2007cs}, which up to its second moment is indistinguishable from a Haar-typical unitary.

Consider decomposing the particular state $\ket{\psi_{b_\star}}_\text{eff}$ according to the two branching structures implied by the two detectors and their decohering dynamics,
\begin{equation}
\begin{aligned}
\ket{\psi_{b_\star}}_\text{eff} &= \sum_i c_i^\prime \ket{s_i}_S \ket{d_i}_{D_\mrm{st}} \ket{\alpha_i}_\mathcal{E} &  \text{(stationary)} \\
\ket{\psi_{b_\star}}_\text{eff} &= \sum_j c_j^{\prime\prime} \ket{s_j}_T \ket{d_j}_{D_\mrm{inf}} \ket{\beta_j}_\mathcal{F} &  \text{(infalling)},
\end{aligned}
 \label{eq:topfac}
\end{equation}
where ``stationary'' and ``infalling'' remind us whether we are expressing the state in the basis of the detector stationary outside or infalling inside the black hole.
Let us focus on the branching structure from the stationary point of view and start writing $\mathcal{E}$ in terms of Hilbert-space subfactors as $\ket{\alpha_i}_{\cal E} = \sum_{kl} \mu^i_{kl} \ket{\phi_k}_A \ket{e_l}_E$. A state $\ket{\phi_k}_A$ describing the stretched horizon can be decomposed into its constituent factors in $D_{\rm inf}$, $T$, and $F$,
\be
\ket{\phi_k}_A = \sum_{abc} U^k_{abc}\ket{s_a}_T \ket{d_b}_{D_{\rm inf}} \ket{f_c}_F, \label{eq:phiK}
\ee
where $U^k_{abc}$ implements a unitary change of basis from $abc$ to $k$. 
Black hole scrambling implies that this unitary is, to a good approximation, Haar typical for a generic choice of basis for $A$ and its constituent Hilbert-space factors. We therefore have
\be 
\label{eq:branch2} \ket{\psi_{b_\star}}_\text{eff} = \sum_i c_i^\prime \ket{s_i}_S \ket{d_i}_{D_\mrm{st}} \sum_{kl} \mu^i_{kl} \left( \sum_{abc} U^{k}_{abc} \ket{s_a}_T \ket{d_b}_{D_\mrm{inf}} \ket{f_c}_F  \right) \ket{e_l}_E .
\ee
Rearranging the sums, we have
\be 
\begin{aligned}
\label{eq:branch3} \ket{\psi_{b_\star}}_\text{eff} &= \sum_{ab} \ket{s_a}_T \ket{d_b}_{D_\mrm{inf}} \sum_c \left[ \sum_i c_i^\prime \left( \sum_l \left[ \sum_k \mu^i_{kl} U^k_{abc} \right] \ket{e_l}_E   \right) \ket{s_i}_S \ket{d_i}_{D_\mrm{st}}  \right] \ket{f_c}_F \\
 & =  \sum_a c_a^{\prime\prime} \ket{s_a}_T \ket{d_a}_{D_{\mrm{inf}}} \ket{\beta_a}_\mathcal{F}.
\end{aligned}
\ee
To recapitulate, in \Eq{eq:branch2}, we wrote each $\ket{\alpha_i}_\mathcal{E}$ in an orthonormal basis for the horizon ($A$) and $E$,
taking each horizon state and expanding it in the pointer-state basis for $T$ and $D_{\rm inf}$, along with some arbitrary basis for $F$.
We can also express $\ket{\psi_{b_\star}}_\text{eff}$ in the pointer basis of the infalling detector, writing it with the branching structure as given in the second line of \Eq{eq:branch3}. Hence, if both the infalling and stationary detector have decohered, it must be that $U^k_{abc} = 0$ if $a \neq b$ so that
\be 
\begin{aligned}
\label{eq:branch5} \ket{\psi_{b_\star}}_\text{eff} & = \sum_{a} \ket{s_a}_T \ket{d_a}_{D_\mrm{inf}} \sum_c \left[ \sum_i c_i^\prime \left( \sum_l \left[ \sum_k \mu^i_{kl} U^k_{aac} \right] \ket{e_l}_E   \right) \ket{s_i}_S \ket{d_i}_{D_\mrm{st}}  \right] \ket{f_c}_F \\
 &\equiv \sum_{a} \ket{s_a}_T \ket{d_a}_{D_\mrm{inf}}  \underbrace{\sum_c \left[ \sum_i c_i^\prime \ket{\tilde e^i_{ac}}_E \ket{s_i}_S \ket{d_i}_{D_\mrm{st}}  \right] \ket{f_c}_F}_{c_a^{\prime\prime} \ket{\beta_a}_\mathcal{F}}
\end{aligned}
\ee
That the horizon scrambles means that the components $U^k_{aac}$ are approximately typical with respect to the Haar measure.

Were we to find that $\sum_k \mu^i_{kl} U^k_{aac} \propto \delta^i_a$, then the sum in \Eq{eq:branch5} would collapse to a single term:
\begin{equation} \label{eq:pathological}
\ket{\psi_{b_\star}}_\text{eff} = \sum_{a} \ket{s_a}_T \ket{d_a}_{D_\mrm{inf}} c_a^\prime \ket{s_a}_S \ket{d_a}_{D_\mrm{st}} \left(\sum_c \ket{\tilde e^a_{ac}}_E \ket{f_c}_F \right) \, .
\end{equation}
Such a situation would be pathological because it would mean that pointer states of the black hole exterior would correlate perfectly with states of the black hole interior, which would mean that they would be stable under the action of their joint environment, i.e., classical and long lived. This would seem to contradict what is believed about black hole fast scrambling. Moreover, such a conspiracy between the matrices $\mu$ and $U$ is implausible since $U$ is Haar typical and furthermore dependent on the detector that we choose.
To see this, note that $U$ describes how the state of the stretched horizon decomposes in the infalling detector's pointer basis, while $\mu$ is independent of the detector properties, simply describing the joint state of the stretched horizon and exterior environment, and has no reason to be correlated with the Haar-typical properties of $U$.

Hence, we have shown that the pointer bases for the interior and exterior observer are not compatible. Specifically, \Eq{eq:branch5} shows that the environment states $\ket{\beta_a}_{\cal F}$ for the infalling detector are given by nontrivial transformations (under $\mu^i_{kl} U^k_{aac}$) of the joint state of the exterior system, detector, and environment, along with the interior environment. Similarly, the environment states associated with the pointer basis for the exterior detector are given by nontrivial transformations of the joint state of the interior system, detector, and environment, along with the exterior environment. 

What this means physically is that it is not possible to isolate a single branch of the wave function, via a natural dynamical decoherence process, that corresponds to a pointer state for the interior and exterior detector simultaneously. 
This property of the global wave function reconciles the complementary points of view of infalling and stationary observers, without requiring the existence of a firewall to preserve unitarity.
Black hole complementarity is therefore implemented in Hilbert space in terms of the relationship between pointer states as defined by different observers across a horizon.

\section{Conclusions}
\label{sec:conclusion}

The information paradox, as sharpened by AMPS, seemingly necessitated modifying a cherished pillar of modern physics in  effective field theory, relativity, or unitarity. 
In this work, we argued using decoherence and pointer bases that such a dramatic conclusion is not directly implied by the ingredients of the AMPS discussion. 
In particular, different components of the AMPS argument apply either globally or on individual branches of the wave function: unitarity applies to the global wave function, while the absence of drama at the horizon is a statement about individual semiclassical branches. 
They can therefore, as far as we can tell, be satisfied simultaneously without violating unitarity, monogamy of entanglement, or any other principles of quantum mechanics.

Since the existence of firewalls would stand in gross violation of our classical intuition, we should judge them to be unlikely unless their absence would require violating an even-more-cherished belief, which we have argued it does not.
Given our best current understanding of quantum mechanics and black hole thermodynamics, there is no reason to insist that an observer falling into a black hole sees anything other than a reason to regret their decision.

\section*{Acknowledgments}
We thank Ahmed Almheiri, Raphael Bousso, William Donnelly, Masahiro Hotta, Cindy Keeler, Yasunori Nomura, Don N. Page, Guillaume Verdon, and Koji Yamaguchi for helpful discussions. This work is supported by the U.S. Department of Energy, Office of Science, Office of High Energy Physics, under Award No. DE-SC0011632.
N.B. is supported by the National Science Foundation, under Grant No. 82248-13067-44-PHPXH.
A.C.-D. is supported by a Beatrice and Sai-Wai Fu Graduate Fellowship in Physics and the Gordon and Betty Moore Foundation through Grant No. 776 to the Caltech Moore Center for Theoretical Cosmology and Physics.
J.P. is supported in part by the Simons Foundation and in part by the Natural Sciences and Engineering Research Council of Canada. G.N.R. is supported by the Miller Institute for Basic Research in Science at the University of California, Berkeley.

\bibliographystyle{utphys-modified}
\bibliography{BHPS}

\providecommand{\href}[2]{#2}\begingroup\raggedright\begin{thebibliography}{10}

\bibitem{Hawking:1974sw}
S.~W. Hawking, ``{Particle creation by black holes},''
  \href{http://dx.doi.org/10.1007/BF02345020}{{\em Commun. Math. Phys.}
  {\bfseries 43} (1975) 199}.

\bibitem{Susskind:1993if}
L.~Susskind, L.~Thorlacius, and J.~Uglum, ``{The stretched horizon and black
  hole complementarity},''
  \href{http://dx.doi.org/10.1103/PhysRevD.48.3743}{{\em Phys. Rev. D}
  {\bfseries 48} (1993) 3743},
  \href{http://arxiv.org/abs/hep-th/9306069}{{\ttfamily arXiv:hep-th/9306069}}.

\bibitem{Polchinski:2016hrw}
J.~Polchinski, \href{http://dx.doi.org/10.1142/9789813149441_0006}{``{The black
  hole information problem},''} in {\em {Proceedings, Theoretical Advanced
  Study Institute in Elementary Particle Physics: New Frontiers in Fields and
  Strings (TASI 2015): Boulder, CO, USA, June 1-26, 2015}}, p.~353.
\newblock 2017.
\newblock
\href{http://arxiv.org/abs/1609.04036}{{\ttfamily arXiv:1609.04036}}.
\newblock

\bibitem{Wallace:2017wzs}
D.~Wallace, ``{Why Black Hole Information Loss is Paradoxical},''
  \href{http://arxiv.org/abs/1710.03783}{{\ttfamily arXiv:1710.03783}}.

\bibitem{Almheiri:2012rt}
A.~Almheiri, D.~Marolf, J.~Polchinski, and J.~Sully, ``{Black holes:
  complementarity or firewalls?},''
  \href{http://dx.doi.org/10.1007/JHEP02(2013)062}{{\em JHEP} {\bfseries 02}
  (2013) 062}, \href{http://arxiv.org/abs/1207.3123}{{\ttfamily
  arXiv:1207.3123}}.

\bibitem{Thorne:1986iy}
K.~S. Thorne, R.~H. Price, and D.~A. Macdonald, eds., {\em {Black holes: the
  membrane paradigm}}.
\newblock Yale University Press,
1986.
\newblock

\bibitem{PhysRevLett.30.434}
E.~H. Lieb and M.~B. Ruskai, ``A fundamental property of quantum-mechanical
  entropy,'' \href{http://dx.doi.org/10.1103/PhysRevLett.30.434}{{\em Phys.
  Rev. Lett.} {\bfseries 30} (1973) 434}.

\bibitem{Almheiri:2013hfa}
A.~Almheiri, D.~Marolf, J.~Polchinski, D.~Stanford, and J.~Sully, ``{An
  apologia for firewalls},''
  \href{http://dx.doi.org/10.1007/JHEP09(2013)018}{{\em JHEP} {\bfseries 09}
  (2013) 018}, \href{http://arxiv.org/abs/1304.6483}{{\ttfamily
  arXiv:1304.6483}}.

\bibitem{Papadodimas:2012aq}
K.~Papadodimas and S.~Raju, ``{An infalling observer in AdS/CFT},''
  \href{http://dx.doi.org/10.1007/JHEP10(2013)212}{{\em JHEP} {\bfseries 10}
  (2013) 212}, \href{http://arxiv.org/abs/1211.6767}{{\ttfamily
  arXiv:1211.6767}}.

\bibitem{Lloyd:2013bza}
S.~Lloyd and J.~Preskill, ``{Unitarity of black hole evaporation in final-state
  projection models},'' \href{http://dx.doi.org/10.1007/JHEP08(2014)126}{{\em
  JHEP} {\bfseries 08} (2014) 126},
  \href{http://arxiv.org/abs/1308.4209}{{\ttfamily arXiv:1308.4209}}.

\bibitem{Unruh:2017uaw}
W.~G. Unruh and R.~M. Wald, ``{Information loss},''
  \href{http://dx.doi.org/10.1088/1361-6633/aa778e}{{\em Rept. Prog. Phys.}
  {\bfseries 80} (2017) 092002},
  \href{http://arxiv.org/abs/1703.02140}{{\ttfamily arXiv:1703.02140}}.

\bibitem{Maldacena:2013xja}
J.~Maldacena and L.~Susskind, ``{Cool horizons for entangled black holes},''
  \href{http://dx.doi.org/10.1002/prop.201300020}{{\em Fortsch. Phys.}
  {\bfseries 61} (2013) 781}, \href{http://arxiv.org/abs/1306.0533}{{\ttfamily
  arXiv:1306.0533}}.

\bibitem{Mathur:2005zp}
S.~D. Mathur, ``{The fuzzball proposal for black holes: an elementary
  review},'' \href{http://dx.doi.org/10.1002/prop.200410203}{{\em Fortsch.
  Phys.} {\bfseries 53} (2005) 793},
  \href{http://arxiv.org/abs/hep-th/0502050}{{\ttfamily arXiv:hep-th/0502050}}.

\bibitem{Nomura:2014woa}
Y.~Nomura, F.~Sanches, and S.~J. Weinberg, ``{Black Hole Interior in Quantum
  Gravity},'' \href{http://dx.doi.org/10.1103/PhysRevLett.114.201301}{{\em
  Phys. Rev. Lett.} {\bfseries 114} (2015) 201301},
  \href{http://arxiv.org/abs/1412.7539}{{\ttfamily arXiv:1412.7539}}.

\bibitem{Hertog:2017vod}
T.~Hertog and J.~Hartle, ``{Observational Implications of Fuzzball
  Formation},''
\href{http://arxiv.org/abs/1704.02123}{{\ttfamily arXiv:1704.02123}}.

\bibitem{Harlow:2013tf}
D.~Harlow and P.~Hayden, ``{Quantum computation vs. firewalls},''
  \href{http://dx.doi.org/10.1007/JHEP06(2013)085}{{\em JHEP} {\bfseries 06}
  (2013) 085}, \href{http://arxiv.org/abs/1301.4504}{{\ttfamily
  arXiv:1301.4504}}.

\bibitem{Bao:2015hdp}
N.~Bao, A.~Bouland, and S.~P. Jordan, ``{Grover Search and the No-Signaling
  Principle},'' \href{http://dx.doi.org/10.1103/PhysRevLett.117.120501}{{\em
  Phys. Rev. Lett.} {\bfseries 117} (2016) 120501},
\href{http://arxiv.org/abs/1511.00657}{{\ttfamily arXiv:1511.00657}}.

\bibitem{Bao:2016uan}
N.~Bao, A.~Bouland, A.~Chatwin-Davies, J.~Pollack, and H.~Yuen, ``{Rescuing
  complementarity with little drama},''
  \href{http://dx.doi.org/10.1007/JHEP12(2016)026}{{\em JHEP} {\bfseries 12}
  (2016) 026}, \href{http://arxiv.org/abs/1607.05141}{{\ttfamily
  arXiv:1607.05141}}.

\bibitem{Chen:2014jwq}
P.~Chen, Y.~C. Ong, and D.-h. Yeom, ``{Black hole remnants and the information
  loss paradox},'' \href{http://dx.doi.org/10.1016/j.physrep.2015.10.007}{{\em
  Phys. Rept.} {\bfseries 603} (2015) 1},
  \href{http://arxiv.org/abs/1412.8366}{{\ttfamily arXiv:1412.8366}}.

\bibitem{Giddings:2012gc}
S.~B. Giddings, ``{Nonviolent nonlocality},''
  \href{http://dx.doi.org/10.1103/PhysRevD.88.064023}{{\em Phys. Rev. D}
  {\bfseries 88} (2013) 064023},
  \href{http://arxiv.org/abs/1211.7070}{{\ttfamily arXiv:1211.7070}}.

\bibitem{Osuga:2016htn}
K.~Osuga and D.~N. Page, ``{Qubit Transport Model for Unitary Black Hole
  Evaporation Without Firewalls},''
  \href{http://arxiv.org/abs/1607.04642}{{\ttfamily arXiv:1607.04642}}.

\bibitem{Hsu:2013cw}
S.~D.~H. Hsu, ``{Macroscopic Superpositions and Black Hole Unitarity},''
  \href{http://arxiv.org/abs/1302.0451}{{\ttfamily arXiv:1302.0451}}.

\bibitem{Hsu:2013fra}
S.~D.~H. Hsu, ``{Factorization of Unitarity and Black Hole Firewalls},''
  \href{http://arxiv.org/abs/1308.5686}{{\ttfamily arXiv:1308.5686}}.

\bibitem{Hartle:2015bna}
J.~Hartle and T.~Hertog, ``{Quantum transitions between classical histories},''
  \href{http://dx.doi.org/10.1103/PhysRevD.92.063509}{{\em Phys. Rev. D}
  {\bfseries 92} (2015) 063509},
\href{http://arxiv.org/abs/1502.06770}{{\ttfamily arXiv:1502.06770}}.

\bibitem{Papadodimas:2015jra}
K.~Papadodimas and S.~Raju, ``{Remarks on the necessity and implications of
  state-dependence in the black hole interior},''
  \href{http://dx.doi.org/10.1103/PhysRevD.93.084049}{{\em Phys. Rev. D}
  {\bfseries 93} (2016) 084049},
  \href{http://arxiv.org/abs/1503.08825}{{\ttfamily arXiv:1503.08825}}.

\bibitem{Nomura:2013gna}
Y.~Nomura, J.~Varela, and S.~J. Weinberg, ``{Black holes or firewalls: a theory
  of horizons},'' \href{http://dx.doi.org/10.1103/PhysRevD.88.084052}{{\em
  Phys. Rev. D} {\bfseries 88} (2013) 084052},
  \href{http://arxiv.org/abs/1308.4121}{{\ttfamily arXiv:1308.4121}}.

\bibitem{Verlinde:2013uja}
E.~Verlinde and H.~Verlinde, ``{Passing Through the Firewall},''
  \href{http://arxiv.org/abs/1306.0515}{{\ttfamily arXiv:1306.0515}}.

\bibitem{Nomura:2012sw}
Y.~Nomura, J.~Varela, and S.~J. Weinberg, ``{Complementarity endures: no
  firewall for an infalling observer},''
  \href{http://dx.doi.org/10.1007/JHEP03(2013)059}{{\em JHEP} {\bfseries 03}
  (2013) 059}, \href{http://arxiv.org/abs/1207.6626}{{\ttfamily
  arXiv:1207.6626}}.

\bibitem{Nomura:2012ex}
Y.~Nomura and J.~Varela, ``{A note on (no) firewalls: the entropy argument},''
  \href{http://dx.doi.org/10.1007/JHEP07(2013)124}{{\em JHEP} {\bfseries 07}
  (2013) 124},
\href{http://arxiv.org/abs/1211.7033}{{\ttfamily arXiv:1211.7033}}.

\bibitem{Yeom:2016qec}
D.-h. Yeom, \href{http://dx.doi.org/10.1142/9789813203952_0079}{``{Information
  loss problem and roles of instantons},''} in {\em {Proceedings, 2nd LeCosPA
  Symposium: Everything about Gravity, Celebrating the Centenary of Einstein's
  General Relativity (LeCosPA2015): Taipei, Taiwan, December 14-18, 2015}},
  p.~566.
\newblock 2017.
\newblock
\href{http://arxiv.org/abs/1601.02366}{{\ttfamily arXiv:1601.02366}}.
\newblock

\bibitem{Bousso:2013wia}
R.~Bousso, ``{Firewalls from double purity},''
  \href{http://dx.doi.org/10.1103/PhysRevD.88.084035}{{\em Phys. Rev. D}
  {\bfseries 88} (2013) 084035},
  \href{http://arxiv.org/abs/1308.2665}{{\ttfamily arXiv:1308.2665}}.

\bibitem{Giddings}
S.~B. Giddings and R.~A. Porto, ``{The gravitational S-matrix},''
  \href{http://dx.doi.org/10.1103/PhysRevD.81.025002}{{\em Phys. Rev. D}
  {\bfseries 81} (2010) 025002},
  \href{http://arxiv.org/abs/0908.0004}{{\ttfamily arXiv:0908.0004}}.

\bibitem{Strominger}
A.~Strominger, ``{On BMS invariance of gravitational scattering},''
  \href{http://dx.doi.org/10.1007/JHEP07(2014)152}{{\em JHEP} {\bfseries 07}
  (2014) 152}, \href{http://arxiv.org/abs/1312.2229}{{\ttfamily
  arXiv:1312.2229}}.

\bibitem{Page:1993wv}
D.~N. Page, ``{Information in Black Hole Radiation},''
  \href{http://dx.doi.org/10.1103/PhysRevLett.71.3743}{{\em Phys. Rev. Lett.}
  {\bfseries 71} (1993) 3743},
  \href{http://arxiv.org/abs/hep-th/9306083}{{\ttfamily arXiv:hep-th/9306083}}.

\bibitem{Harlow:2014yka}
D.~Harlow, ``{Jerusalem lectures on black holes and quantum information},''
  \href{http://dx.doi.org/10.1103/RevModPhys.88.015002}{{\em Rev. Mod. Phys.}
  {\bfseries 88} (2016) 015002},
  \href{http://arxiv.org/abs/1409.1231}{{\ttfamily arXiv:1409.1231}}.

\bibitem{Maudlin:2017lye}
T.~Maudlin, ``{(Information) Paradox Lost},''
\href{http://arxiv.org/abs/1705.03541}{{\ttfamily arXiv:1705.03541}}.

\bibitem{Fitzpatrick:2011dm}
A.~L. Fitzpatrick and J.~Kaplan, ``{Unitarity and the holographic S-matrix},''
  \href{http://dx.doi.org/10.1007/JHEP10(2012)032}{{\em JHEP} {\bfseries 10}
  (2012) 032}, \href{http://arxiv.org/abs/1112.4845}{{\ttfamily
  arXiv:1112.4845}}.

\bibitem{1967PhRv..160.1113D}
B.~S. {Dewitt}, ``{Quantum theory of gravity. I. the canonical theory},''
  \href{http://dx.doi.org/10.1103/PhysRev.160.1113}{{\em Phys. Rev.} {\bfseries
  160} (1967) 1113}.

\bibitem{Zeh:1970fop}
H.~D. {Zeh}, ``{On the interpretation of measurement in quantum theory},''
  \href{http://dx.doi.org/10.1007/BF00708656}{{\em Found. Phys.} {\bfseries 1}
  (1970) 69}.

\bibitem{Zurek:1981xq}
W.~H. Zurek, ``Pointer basis of quantum apparatus: into what mixture does the
  wave packet collapse?,''
  \href{http://dx.doi.org/10.1103/PhysRevD.24.1516}{{\em Phys. Rev. D}
  {\bfseries 24} (1981) 1516}.

\bibitem{Griffiths:1984rx}
R.~B. Griffiths, ``{Consistent histories and the interpretation of quantum
  mechanics},'' \href{http://dx.doi.org/10.1007/BF01015734}{{\em J. Statist.
  Phys.} {\bfseries 36} (1984) 219}.

\bibitem{Joos:1984uk}
E.~Joos and H.~D. Zeh, ``{The emergence of classical properties through
  interaction with the environment},''
  \href{http://dx.doi.org/10.1007/BF01725541}{{\em Z. Phys. B} {\bfseries 59}
  (1985) 223}.

\bibitem{Schlosshauer:2003zy}
M.~Schlosshauer, ``{Decoherence, the measurement problem, and interpretations
  of quantum mechanics},''
  \href{http://dx.doi.org/10.1103/RevModPhys.76.1267}{{\em Rev. Mod. Phys.}
  {\bfseries 76} (2004) 1267},
  \href{http://arxiv.org/abs/quant-ph/0312059}{{\ttfamily
  arXiv:quant-ph/0312059}}.

\bibitem{Jacobson:2015hqa}
T.~Jacobson, ``{Entanglement Equilibrium and the Einstein Equation},''
  \href{http://dx.doi.org/10.1103/PhysRevLett.116.201101}{{\em Phys. Rev.
  Lett.} {\bfseries 116} (2016) 201101},
\href{http://arxiv.org/abs/1505.04753}{{\ttfamily arXiv:1505.04753}}.

\bibitem{Harlow:2016vwg}
D.~Harlow, ``{The Ryu-Takayanagi formula from quantum error correction},''
  \href{http://dx.doi.org/10.1007/s00220-017-2904-z}{{\em Commun. Math. Phys.}
  {\bfseries 354} (2017) 865},
\href{http://arxiv.org/abs/1607.03901}{{\ttfamily arXiv:1607.03901}}.

\bibitem{Cao:2017hrv}
C.~Cao and S.~M. Carroll, ``{Bulk Entanglement Gravity without a Boundary:
  Towards Finding Einstein's Equation in Hilbert Space},''
\href{http://arxiv.org/abs/1712.02803}{{\ttfamily arXiv:1712.02803}}.

\bibitem{Carney:2017jut}
D.~Carney, L.~Chaurette, D.~Neuenfeld, and G.~W. Semenoff, ``{Infrared quantum
  information},'' \href{http://dx.doi.org/10.1103/PhysRevLett.119.180502}{{\em
  Phys. Rev. Lett.} {\bfseries 119} no.~18, (2017) 180502},
\href{http://arxiv.org/abs/1706.03782}{{\ttfamily arXiv:1706.03782}}.

\bibitem{Carney:2017oxp}
D.~Carney, L.~Chaurette, D.~Neuenfeld, and G.~W. Semenoff, ``{Dressed infrared
  quantum information},''
  \href{http://dx.doi.org/10.1103/PhysRevD.97.025007}{{\em Phys. Rev.}
  {\bfseries D97} no.~2, (2018) 025007},
\href{http://arxiv.org/abs/1710.02531}{{\ttfamily arXiv:1710.02531}}.

\bibitem{Carney:2018ygh}
D.~Carney, L.~Chaurette, D.~Neuenfeld, and G.~Semenoff, ``{On the need for soft
  dressing},''
\href{http://arxiv.org/abs/1803.02370}{{\ttfamily arXiv:1803.02370}}.

\bibitem{Strominger:2014pwa}
A.~Strominger and A.~Zhiboedov, ``{Gravitational Memory, BMS Supertranslations
  and Soft Theorems},'' \href{http://dx.doi.org/10.1007/JHEP01(2016)086}{{\em
  JHEP} {\bfseries 01} (2016) 086},
\href{http://arxiv.org/abs/1411.5745}{{\ttfamily arXiv:1411.5745}}.

\bibitem{Hawking:2016msc}
S.~W. Hawking, M.~J. Perry, and A.~Strominger, ``{Soft Hair on Black Holes},''
  \href{http://dx.doi.org/10.1103/PhysRevLett.116.231301}{{\em Phys. Rev.
  Lett.} {\bfseries 116} no.~23, (2016) 231301},
\href{http://arxiv.org/abs/1601.00921}{{\ttfamily arXiv:1601.00921}}.

\bibitem{Strominger:2017aeh}
A.~Strominger, ``{Black Hole Information Revisited},''
\href{http://arxiv.org/abs/1706.07143}{{\ttfamily arXiv:1706.07143 [hep-th]}}.

\bibitem{Marolf:2012xe}
D.~Marolf and A.~C. Wall, ``{Eternal Black Holes and Superselection in
  AdS/CFT},'' \href{http://dx.doi.org/10.1088/0264-9381/30/2/025001}{{\em
  Class. Quant. Grav.} {\bfseries 30} (2013) 025001},
\href{http://arxiv.org/abs/1210.3590}{{\ttfamily arXiv:1210.3590}}.

\bibitem{Nomura:2011dt}
Y.~Nomura, ``{Physical Theories, Eternal Inflation, and Quantum Universe},''
  \href{http://dx.doi.org/10.1007/JHEP11(2011)063}{{\em JHEP} {\bfseries 11}
  (2011) 063},
\href{http://arxiv.org/abs/1104.2324}{{\ttfamily arXiv:1104.2324 [hep-th]}}.

\bibitem{Pollack:2018yum}
J.~Pollack and A.~Singh, ``{Towards Space from Hilbert Space: Finding Lattice
  Structure in Finite-Dimensional Quantum Systems},''
\href{http://arxiv.org/abs/1801.10168}{{\ttfamily arXiv:1801.10168}}.

\bibitem{Holevo98}
A.~S. Holevo, ``The capacity of the quantum channel with general signal
  states,'' \href{http://dx.doi.org/10.1109/18.651037}{{\em {IEEE} Trans. Inf.
  Theory} {\bfseries 44} (1998) 269},
  \href{http://arxiv.org/abs/quant-ph/9611023}{{\ttfamily
  arXiv:quant-ph/9611023}}.

\bibitem{Bao:2017guc}
N.~Bao and H.~Ooguri, ``{Distinguishability of black hole microstates},''
  \href{http://dx.doi.org/10.1103/PhysRevD.96.066017}{{\em Phys. Rev. D}
  {\bfseries 96} (2017) 066017},
  \href{http://arxiv.org/abs/1705.07943}{{\ttfamily arXiv:1705.07943}}.

\bibitem{Mathur:2009zs}
S.~D. Mathur, ``{How fast can a black hole release its information?},''
  \href{http://dx.doi.org/10.1142/S0218271809016004}{{\em Int. J. Mod. Phys. D}
  {\bfseries 18} (2009) 2215}, \href{http://arxiv.org/abs/0905.4483}{{\ttfamily
  arXiv:0905.4483}}.

\bibitem{Sekino:2008he}
Y.~Sekino and L.~Susskind, ``{Fast scramblers},''
  \href{http://dx.doi.org/10.1088/1126-6708/2008/10/065}{{\em JHEP} {\bfseries
  10} (2008) 065}, \href{http://arxiv.org/abs/0808.2096}{{\ttfamily
  arXiv:0808.2096}}.

\bibitem{Hayden:2007cs}
P.~Hayden and J.~Preskill, ``{Black holes as mirrors: quantum information in
  random subsystems},''
  \href{http://dx.doi.org/10.1088/1126-6708/2007/09/120}{{\em JHEP} {\bfseries
  09} (2007) 120}, \href{http://arxiv.org/abs/0708.4025}{{\ttfamily
  arXiv:0708.4025}}.

\end{thebibliography}\endgroup

\end{document}